\documentclass[sigconf]{acmart}
\renewcommand\footnotetextcopyrightpermission[1]{} 
\AtBeginDocument{%
  \providecommand\BibTeX{{%
    \normalfont B\kern-0.5em{\scshape i\kern-0.25em b}\kern-0.8em\TeX}}}

\begin{document}


\title{The Last Decade in Review: Tracing the Evolution of Safety Assurance Cases through a Comprehensive Bibliometric Analysis}

\author{Mithila Sivakumar}
\affiliation{%
  \institution{York University}
  \streetaddress{4700 Keele St}
  \city{Toronto}
  \country{Canada}}
\email{msivakum@yorku.ca}

\author{Alvine Boaye Belle}
\affiliation{%
  \institution{York University}
  \streetaddress{4700 Keele St}
  \city{Toronto}
  \country{Canada}}
\email{alvine.belle@lassonde.yorku.ca}

\author{Jinjun Shan}
\affiliation{%
  \institution{York University}
  \streetaddress{4700 Keele St}
  \city{Toronto}
  \country{Canada}}
\email{jjshan@yorku.ca}

\author{Opeyemi Adesina}
\affiliation{%
  \institution{University of the Fraser Valley}
  \streetaddress{33844 King Rd}
  \city{Abbotsford}
  \country{Canada}}
\email{opeyemi.adesina@ufv.ca}

\author{Song Wang}
\affiliation{%
  \institution{York University}
  \streetaddress{4700 Keele St}
  \city{Toronto}
  \country{Canada}}
\email{wangsong@yorku.ca}

\author{Marsha Chechik}
\affiliation{%
  \institution{University of Toronto}
  \streetaddress{27 King's College Cir}
  \city{Toronto}
  \country{Canada}}
\email{chechik@cs.toronto.edu}

\author{Marios Fokaefs}
\affiliation{%
  \institution{York University}
  \streetaddress{4700 Keele St}
  \city{Toronto}
  \country{Canada}}
\email{fokaefs@yorku.ca}

\author{Kimya Khakzad Shahandashti}
\affiliation{%
  \institution{York University}
  \streetaddress{4700 Keele St}
  \city{Toronto}
  \country{Canada}}
\email{kimya@yorku.ca}

\author{Oluwafemi Odu}
\affiliation{%
  \institution{York University}
  \streetaddress{4700 Keele St}
  \city{Toronto}
  \country{Canada}}
\email{olufemi2@yorku.ca}


\begin{abstract}

Safety assurance is of paramount importance across various domains, including automotive, aerospace, and nuclear energy, where the reliability and acceptability of mission-critical systems are imperative. This assurance is effectively realized through the utilization of Safety Assurance Cases. The use of safety assurance cases allows for verifying the correctness of the created systems’ capabilities, preventing system failure. The latter may result in loss of life, severe injuries, large-scale environmental damage, property destruction, and major economic loss. Still, the emergence of complex technologies such as Cyber-Physical Systems (CPSs), characterized by their heterogeneity, autonomy, machine learning capabilities, and the uncertainty of their operational environments poses significant challenges for safety assurance activities. Several papers have tried to propose solutions to tackle these challenges, but to the best of our knowledge, no secondary study investigates the trends, patterns, and relationships characterizing the safety case scientific literature. This makes it difficult to have a holistic view of the safety case landscape and to identify the most promising future research directions. In this paper, we, therefore, rely on state-of-the-art bibliometric tools (e.g., VosViewer) to conduct a bibliometric analysis that allows us to generate valuable insights, identify key authors and venues, and gain a bird’s eye view of the current state of research in the safety assurance area. By revealing knowledge gaps and highlighting potential avenues for future research, our analysis provides an essential foundation for researchers, corporate safety analysts, and regulators seeking to embrace or enhance safety practices that align with their specific needs and objectives.
\end{abstract}



\maketitle
\pagestyle{empty} 

\section{Introduction}
Safety incidents involving autonomous vehicles and other safety-critical systems can lead to fatal outcomes and pose significant risks to manufacturers and the public \cite{b1}. Recent tragedies, like the implosion of the Titan submersible \cite{b57}, highlight the need for prioritizing safety and complying with industry standards. In order to prevent such disasters, safety-critical systems (e.g., cyber-physical systems) are subject to regulation through domain-specific safety standards like ISO 26262, IEC 61508, DO-178C \cite{b2, b10, b17, b54, b58}. To certify these systems, regulatory authorities have established different means by which the most common practice documents safety assurance cases (SACs). Enforcing mandatory safety measures and utilizing SACs— also called safety cases — can prevent such accidents and safeguard lives. Safety assurance is essential across several industries such as nuclear energy, aerospace, and transportation. It usually involves building a convincing argument that aims at demonstrating systems are safe enough to operate in specific environments. Thus, the use of safety cases allows verifying the correct implementation of the desired systems’ properties (or requirements). This allows preventing the failure of such systems. That failure may result in loss of life, severe injuries, large-scale environmental damage, property destruction, and major economic and capital loss \cite{b16}.  

A powerful approach for applying a set of quantitative metrics to evaluate the influence of academic contributions within a specific field is bibliometrics \cite{b3}. Bibliometrics is widely recognized as an effective method for uncovering the cumulative knowledge structures and intellectual associations within a specific field or discipline \cite{b3, b18}. Several bibliometric analyses (e.g., \cite{b3, b9, b19, b20}) have therefore been performed in various domains (e.g., energy, information technology, finance) and have received significant attention. Still, even though there is a rich body of literature on SACs, there is — to the best of our knowledge — no secondary study that has performed a detailed bibliometric analysis to investigate the trends, patterns, and relationships characterizing that literature. This makes it challenging to get a holistic view of the SAC landscape and hinders the identification of the most promising future research directions in that area. 

Therefore, we conduct a bibliometric analysis on the subject of safety assurance cases with the purpose of understanding its overall intellectual landscape. To achieve this, we utilize descriptive statistics to analyze primary studies, authors, venues, and other relevant information gathered from the literature. The synthesis provides valuable insights into the characteristics and dynamics of safety case research, contributing to a comprehensive understanding of the field's current state and outlining potential future research directions. For this study, we selected 224 primary studies relevant to SACs, from the period spanning 2012 to 2022. We relied on five well-established databases to select these studies.

The bibliometric analysis we report in this paper will be of great assistance to researchers and practitioners (e.g., corporate safety analysts, and regulators) involved in the development and certification of mission-critical systems. More specifically, by providing valuable insights, trends, and patterns in the literature on SACs, this analysis will guide them in exploring, implementing, and advancing safety assurance solutions. Our findings will help the audience understand the current state of safety case research and identify potential areas for improvement and innovation in safety assurance practices. 

The remainder of this paper is organized as follows. Section \ref{sec2} explains some background concepts. Section \ref{sec3} describes the methodology we used to perform our bibliometric analysis. Section \ref{sec4} presents the results of that analysis. Section \ref{sec6} reports the limitations of the analysis, and finally Section \ref{sec7} concludes and outlines future work. 

\section{Background and Related work} \label{sec2}

\subsection{Background}
An assurance case is a “\textit{set of audible claims, arguments, and evidence created to support the claim that a defined system/service will satisfy particular requirements}” \cite{b59, b60}. An assurance case is a document that eases the exchange of information between various system stakeholders (e.g., suppliers, acquirers), and between the operator and regulator, where the knowledge regarding a system’s requirements (e.g., safety, security, reliability) is convincingly conveyed \cite{b60}. Assurance cases are structured as a hierarchy of claims, with lower-level claims drawing on concrete evidence, and serving as evidence to justify claims higher in the hierarchy \cite{b62, b63}. Several standards (e.g., ISO 26262) advocate the use of assurance cases to support the certification of various systems, including cyber-physical systems (CPSs) such as autonomous systems \cite{b62, b64}. Depending on the target requirement, there are several types of assurance cases, e.g., security cases, safety cases, dependability cases \cite{b15, b81, b60}. Safety cases are assurance cases that support the safety assurance process \cite{b4}. In this study, we focus on safety cases as safety is a life-critical requirement.

A SAC aims to construct a convincing and valid argument, demonstrating that a system is adequately safe for operation within a specific environment. SACs have been employed for several years and are witnessing a growing trend in constructing safety arguments for safety-critical systems across various industries. Domains such as automotive, aerospace, nuclear energy, and healthcare have increasingly adopted SACs as a means to demonstrate and justify the safety of their systems \cite{b5}. This surge in adoption highlights the importance of SACs in assuring the safety of mission-critical systems, ensuring compliance with industry standards (e.g., ISO 26262, DO-178C, UL 4600), and mitigating potential safety risks. 

Safety-critical systems (e.g., CPSs) usually operate in dynamic, complex, and sometimes unpredictable environments. The uncertainty they face during their operations can stem from various sources, including fluctuating environmental conditions or unexpected system behavior. That uncertainty introduces inherent risks that can compromise the safety and reliability of these systems. The use of SACs endeavors to address and mitigate such risks, aiming to establish a robust framework that ensures the system's ability to operate safely, even in the face of uncertainty.

\begin{figure}[t!]
  \centering
  \includegraphics[width=\linewidth]{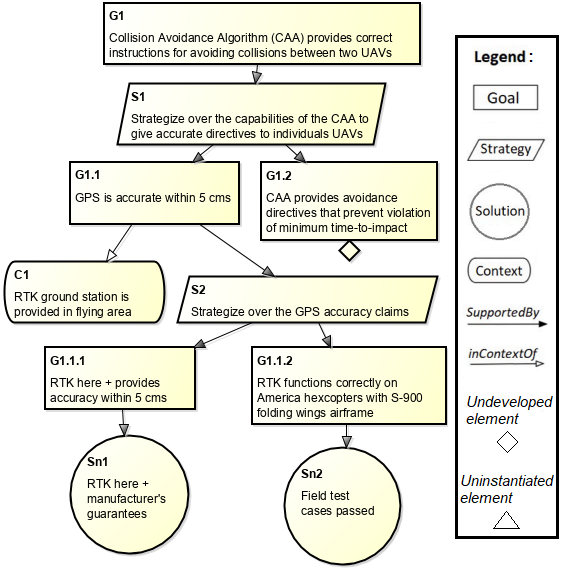}
  \caption{An example of a partial SAC (adapted from \cite{b6}).}
  \Description{This figure shows a sample of a safety case represented using GSN adopted from \cite{b6}. This safety case explains how the top\textcolor{purple}{-}level goal of the correctness of the collision avoidance algorithm is achieved through a series of sub-goals, contexts, and strategies.}
  \label{fig1}
\end{figure}


Several textual and graphical notations allow representing SACs \cite{b5, b71}. Graphical notations include CAE (Claim-Argument-Evidence) \cite{b74} and the very popular GSN (Goal Structuring Notation) \cite{b75}. GSN represents a safety case as a tree-like structure called \textit{goal structure} \cite{b76}. The GSN working group further describes the GSN specification \cite{b75}. GSN core elements include goals, strategies, solutions, and contexts \cite{b77}. A GSN goal depicts a claim. A GSN strategy depicts an argument and embodies the inference rules that allow inferring a claim from sub-claims. A GSN solution depicts a piece of evidence. A GSN element can be decorated using the following decorators, i.e., \textit{uninstantiated} and \textit{undeveloped} \cite{b84}. Two relationships can be used to connect GSN elements: \textit{SupportedBy} and \textit{InContextOf} \cite{b83}. GSN and CAE are aligned with the SACM (Structured Assurance Case Metamodel) \cite{b59} standard that the OMG (Object Management Group) issued and specified to promote standardization and interoperability in assurance case development \cite{b78}. 

Figure \ref{fig1} is adapted from \cite{b6}. It depicts a partial safety case represented using the GSN. That safety case focuses on demonstrating that the collision avoidance algorithm used in the unmanned aerial vehicle (UAV) at hand is able to correctly avoid collision.

\subsection{Related Work}
Bibliometric analysis has been adopted in several different domains including medical, and energy. For example, Han et al. \cite{b93} conducted a comprehensive bibliometric analysis of research trends in surgery with mixed reality from 2000 to 2019, revealing key areas of focus such as training tools, clinical applications, etc. 
Another study by Qin et al. \cite{b3} conducted a comprehensive bibliometric review of past green energy adoption (GEA) research and its determinants. It provided insights into authors, countries, journals, and the evolution of GEA research. 

In the remainder of this section, we discuss existing secondary studies (i.e., \cite{b7, b8, b9, b15, b10}) related to our study but with a focus on system assurance (e.g., safety assurance, security assurance).

In the field of safety assurance, Nair et al. \cite{b7} developed a taxonomy for the classifications of information and artifacts considered as safety evidence in various application domains. Tambon et al. \cite{b8} identified and analyzed challenges associated with certifying machine learning-based systems and explored proposed solutions found in the literature. Neto et al. \cite{b9} provided an overview of the state-of-the-art in the safety assurance of Artificial Intelligence (AI)-based systems and offered guidelines for future work. Mohamad et al. \cite{b15}, focused on a different type of assurance cases called security cases. They therefore conducted a systematic literature review on Security Assurance Cases to highlight the need for validated methodologies, practical guidance, and dedicated tools to support security case development.  Maksimov et al. \cite{b10} examined the assessment features implemented in 10 assurance case software tools, discussed their strengths and weaknesses, and identified future research directions.

None of the aforementioned studies offers an advanced bibliometric synthesis of the existing body of literature on SACs. Most of them mainly provide individual perspectives on specific aspects of safety assurance, such as taxonomy development, AI and machine-learning-based systems, and assurance case tools. Still, they do not comprehensively analyze the entire scientific landscape of safety assurance case research through a bibliometric lens. Our research aims to address this gap by synthesizing collective knowledge within the field of safety cases, providing a holistic perspective that complements the insights gained from these individual studies.

\section{Methodology} \label{sec3}

\begin{figure}[t!]
  \centering
  \includegraphics[width=\linewidth]{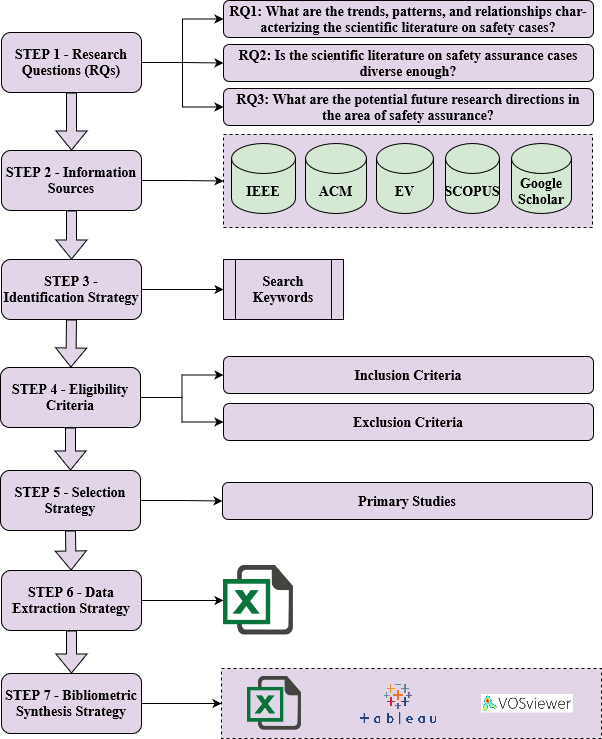}
  \caption{Overview of our bibliometric methodology.}
  \Description{This figure showcases our bibliometric analysis framework by explaining the steps involved in our methodology.}
  \label{fig2}
\end{figure}

The methodology we have adopted to conduct our bibliometric analysis builds on the features of bibliometric analysis guidelines \cite{b65}. These guidelines recommend using several techniques (e.g., performance analysis and science mapping techniques) to quantitatively review the literature on a research topic. Our methodology also builds on several features from well-established reporting guidelines used to conduct secondary studies. These include PRISMA (Preferred Reporting Items for Systematic Reviews and Meta-Analyses) 2020 \cite{b40} and SEGRESS (Software Engineering Guidelines for Reporting Secondary Studies) \cite{b41}. Figure \ref{fig2} provides an overview of the methodology we employed for our bibliometric analysis. We further describe our methodology in the remainder of this section.
    
\subsection{Step 1: definition of research questions} \label{sec32}

The main goal of our bibliometric analysis is to provide some descriptive statistics to identify trends, patterns, and relationships characterizing the scientific literature on SACs. The outcomes of our analysis are of significant importance, as they help identify gaps in safety assurance practices while highlighting areas of expertise and identifying potential directions for future research. 

To achieve that goal, we investigate the following research questions (RQs) when conducting our bibliometric analysis:  

\noindent \textbf{RQ1: What are the trends, patterns, and relationships characterizing the scientific literature on safety cases?}
In this research question, we aim to uncover the evolving landscape of safety case research by analyzing the publication years and venues of the selected primary studies. This is crucial for gaining insights into the trajectory of the field.

\noindent \textbf{RQ2: Is the scientific literature on safety assurance cases diverse enough?}
In this question, we aim to explore the intricate web of authorship within the domain of safety case research under the diversity lens. Examining the patterns and trends among authors not only provides insights into the composition of the research community but also reveals valuable information about the dynamics of knowledge creation and dissemination.

\noindent \textbf{RQ3: What are the potential future research directions in the area of safety assurance?} 
In this question, we aim to uncover evolving challenges and opportunities in the context of safety assurance, particularly in the rapidly advancing domains of technology such as artificial intelligence, machine learning, and autonomous systems. The answer to this research question will help researchers identify and prioritize critical topics, methodologies, and standards that will shape the future of safety assurance, ultimately contributing to the development of safer systems.

\begin{table*}[t!]
  \caption{Inclusion and exclusion criteria}
  \label{tab:elig}
  \begin{tabular}{ll} 
    \toprule
    Inclusion criteria&Exclusion criteria\\
    \midrule
    1. Peer-reviewed Conference, Journals, Workshops papers  & 1. Books, Posters, Book chapters, Theses, Tutorials, Secondary studies \\
    2. Focus on safety case & 2. Studies that focus on general assurance and safety \\
    3. Publication year between 2012 and 2022 & 3. Non-English publications \\
    & 4. Short papers less than 4 pages\\
    & 5. Studies inaccessible due to paywalls\\
    & 6. Unsuccessful attempts to secure free access from authors \\
    & 7. Studies that do not propose techniques focusing on safety case \\
  \bottomrule
\end{tabular}
\end{table*}
\subsection{Step 2: selection of information sources} \label{sec33}

The most commonly utilized search technique is database-driven, and we employed this approach to search for primary studies. Our search focused on five common databases: IEEE Explore \cite{b85}, ACM Digital Library \cite{b86}, SCOPUS \cite{b87}, Engineering Village (EV) \cite{b88}, and Google Scholar \cite{b89}. For our search on Google Scholar, we relied on the Publish Or Perish tool developed by Harzing \cite{b14}.

\subsection{Step 3: identification strategy specification}\label{sec35}

To identify studies within the databases, we used two queries:

\smallskip
\noindent\fbox{%
    \parbox{\linewidth}{%

\textbf{QUERY-1: }
\textit{"Safety Claim" \textcolor{red}{OR} "Safety argument" \textcolor{red}{OR} "Safety evidence" \textcolor{red}{OR} "Safety justification" \textcolor{red}{OR} "Safety case" \textcolor{red}{OR} “Safety assurance case” \textcolor{red}{OR} "Safety assurance" \textcolor{red}{OR} "Safety compliance" \textcolor{red}{OR} “Safety artifacts”}

\smallskip

\textbf{QUERY-2: }
\textit{("Safety Case") \textcolor{red}{AND} ("GSN" \textcolor{red}{OR} "CAE" \textcolor{red}{OR} "SACM")}

}
}

\smallskip
To ensure the queries used in the database-driven search effectively capture a substantial yet manageable body of relevant literature, two researchers (i.e. a graduate student and a faculty member) from our team meticulously refined the queries multiple times. A third one (i.e., a faculty member) reviewed the query and provided recommendations to make sure it is inclusive enough.
When conducting the database-driven search, we executed the above two search strings in the titles, abstracts, keywords, or metadata of the studies available in the five databases (see Section \ref{sec33}). 

\subsection{Step 4: specification of eligibility criteria}

Table \ref{tab:elig} reports the eligibility criteria we used to select the primary studies we included in our bibliometric analysis.


\subsection{Step 5: specification of the selection strategy}\label{sec39}
To select and filter out primary studies according to the eligibility criteria, we employed a five-phase process explained as follows:
\begin{itemize}
\item \textbf{\textit{Phase-1:}} In this phase, we gathered the records collected from all 5 databases (using both queries) and import them to EndNote, a reference management tool \cite{b47, b11}.
\item \textbf{\textit{Phase-2:}} In this phase, we harnessed the duplicate records detection feature of EndNote tool to eliminate duplicates.
\item \textbf{\textit{Phase-3:}} From the third phase, we manually filtered the records, starting with scanning the records titles based on the inclusion and exclusion criteria 
\item \textbf{\textit{Phase-4:}} In the fourth phase, we scanned the abstracts and titles of the records based on the inclusion and exclusion criteria.
\item \textbf{\textit{Phase-5:}} Finally, we scanned the entire text of the document based on the inclusion and exclusion criteria. This allowed us to get the final list of primary studies on which our bibliometric analysis focuses.
\end{itemize}

A single researcher completed the aforementioned phases to select primary studies. Still, to minimize bias during the selection process, a second researcher randomly selected sampled studies in each phase to validate their selection. Meetings were held to resolve potential disagreements between the two researchers. 

\subsection{Step 6: data extraction strategy formulation}

To collect data from the primary studies, we employed structured data extraction forms designed as Excel sheets. We used these forms to record bibliometric information (e.g., authors, titles, and publication venues) for all the studies incorporated into our study. Furthermore, we utilized the Endnote reference management tool to extract bibliometric data in the form of Research Information Systems (RIS) files, which is needed as input for VOSviewer \cite{b13}.

\subsection{Step 7: specification of the synthesis strategy}
The adoption of bibliometric synthesis is on the rise across various research fields (\cite{b42,b43,b44,b45,b46}). To complete our bibliometric synthesis, we employ the widely used bibliometric tool VosViewer \cite{b13} to extract and analyze bibliometric data from primary studies. This tool facilitates the automated creation of charts presenting the bibliometric characteristics of the primary studies. Additionally, for generating supplementary charts,
we utilize Google Charts \cite{b82}, Microsoft Excel, and leverage Tableau \cite{b12}. The latter is a prominent data visualization tool extensively used in the data analytics sector. 

\section{Results} \label{sec4}

\begin{figure}[t!]
  \centering
  \includegraphics[width=80 mm]{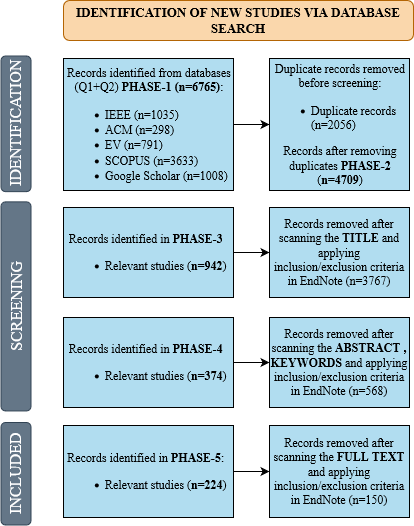}
  \caption{PRISMA Flow diagram of the selection process.}
  \Description{This flow diagram explains the five phase process of filtering the records obtained from the five databases mentioned in Section \ref{sec33} and the number of primary studies obtained at the end of the fifth phase.}
  \label{fig3}
\end{figure}

The PRISMA flow diagram that Figure \ref{fig3} depicts shows the results of each of the six phases that we completed to select the primary studies we included in our bibliometric analysis.
That PRISMA flow diagram comprises three main stages (i.e. \textit{Identification}, \textit{Screening}, and \textit{Included}) that map to the five phases of our selection process. Hence, at the \textit{Identification} stage, we started with 6,765 records that we identified from five databases and imported to EndNote. 
Finally, at the \textit{Included} stage, we selected 224 primary studies that we included in our bibliometric analysis. The list of these studies is available online\footnote{Github link listing all the primary studies included in this study: \url{https://github.com/AnonymousAuthours/SACs-A-bibliometric-Synthesis_SEAMS-2024}}. We rely on these studies to answer our research questions (RQs) (see Section \ref{sec32}) through our bibliometric lens. 

\subsection{RQ1: SAC Scientific Landscape}
\subsubsection{Characteristics of primary studies by year:}

\begin{figure}[t]
  \centering
  \includegraphics[width=\linewidth]{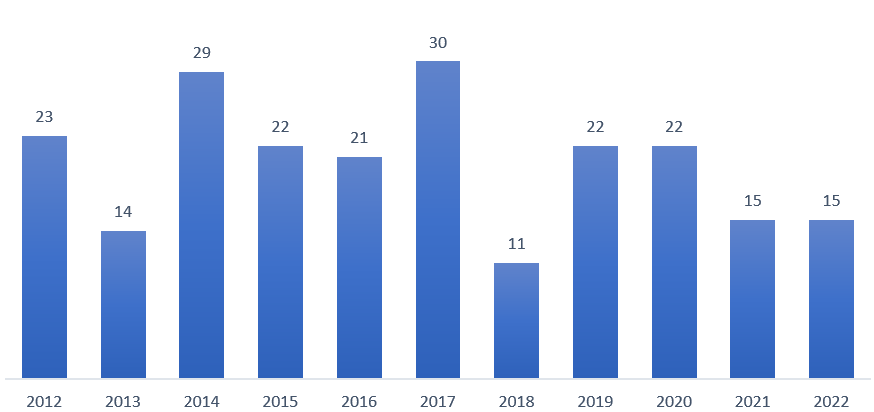}
  \caption{Distribution of primary studies per year (2012-2022)}
  \Description{This bar chart shows the distribution of primary studies between 2012 and 2022.}
  \label{fig4}
\end{figure}

Figure \ref{fig4} illustrates the distribution of primary studies published between 2012 and 2022. This reveals some fluctuations in the number of studies published during that period.

A notable observation drawn from Figure \ref{fig4} is the distinct peak in 2017, with a substantial increase in publications. In contrast to the preceding year, which witnessed the publication of only 21 studies, 2017 marked a significant surge with 30 publications. Similarly, there was a surge in publications in 2019 and 2020, with 22 studies being released, compared to 2018 with only 13 studies being published. Despite the decrease in publications during certain years, such as 2013, 2018, and 2021, the subsequent years show an upward trend, indicating a sustained interest in the topic of SACs. The decrease in 2021 may be due to the COVID-19 pandemic that disrupted research activities across the world and therefore had an adverse impact on the number of publications that year and the subsequent year. Still, with the recent rise of autonomous systems (e.g., autonomous driving vehicles, drones) that need to demonstrate sufficient safe autonomy before being trusted and deployed, we are confident that the research on SACs will remain critical and attractive and will lead to many publications in the upcoming years.

\subsubsection{Characteristics of primary studies by most influential keywords:}
 \label{keywordsection}

\begin{figure*}[h]
  \centering
  \includegraphics[width=\linewidth]{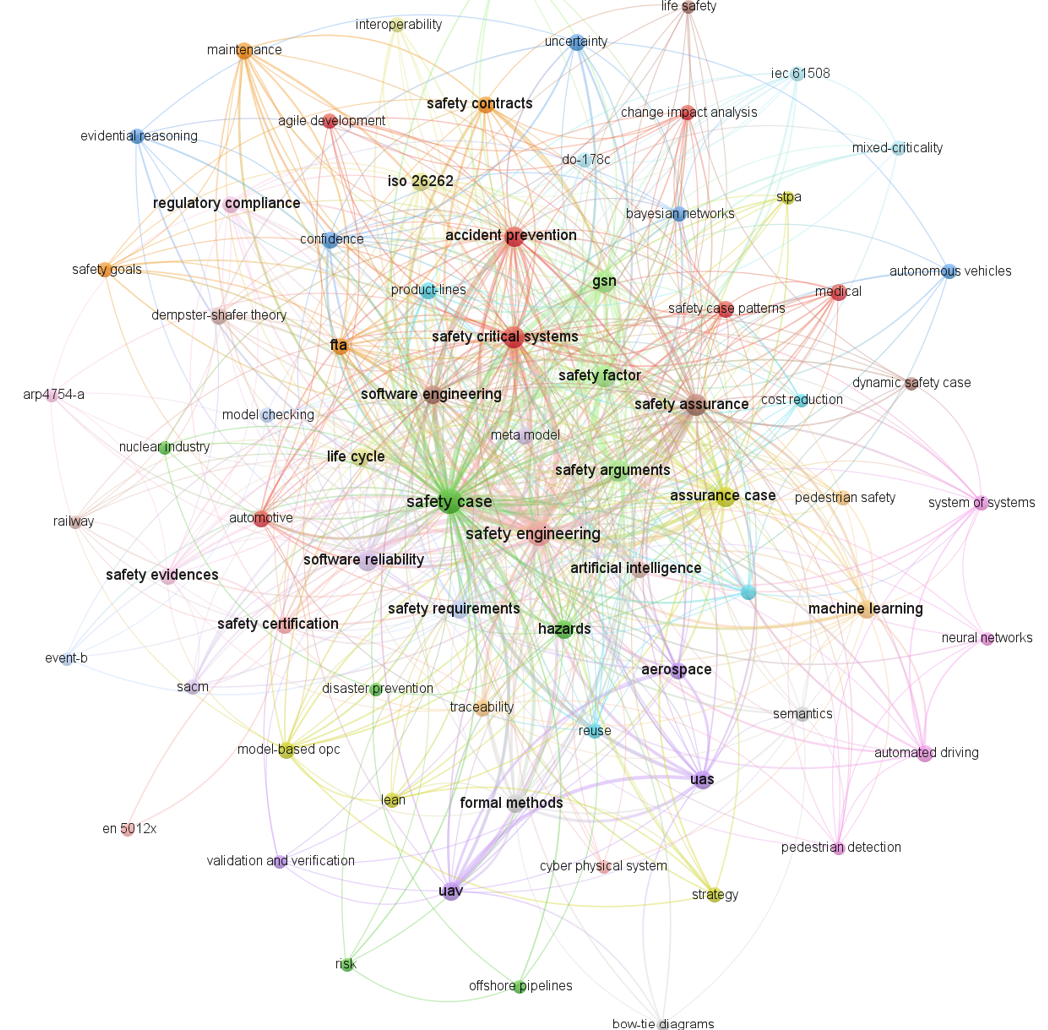}
  \caption{Co-occurence map of most influential keywords.}
  \Description{This map generated from the VOSviewer shows the co-occurence network of the most influential keywords from the 224 primary studies included in this analysis.}
  \label{fig5}
\end{figure*}

We relied on VOSViewer to generate the map that Figure \ref{fig5} displays. That figure illustrates the most frequent keywords occurring in the 224 studies included in this analysis. To create the map while encompassing a wide range of related keywords, in VOSviewer, we set the minimum occurrence limit for keywords to 2. The analysis of the map reveals several interesting findings. First and foremost, the keywords \textbf{\textit{"Safety Case", "Safety Engineering", "Safety Factor", "Safety Arguments", "Safety Case Patterns", "GSN"}} and \textbf{\textit{"Safety Goals"}} emerge as the top terms appearing in a significant number of studies. This aligns with the search queries employed during the database-driven search, which specifically targeted literature related to these terms. The prevalence of these keywords indicates that the included studies indeed directly address the topic of safety cases and safety assurance.

The links between the keywords \textbf{\textit{"Machine Learning", "Artificial Intelligence", "Autonomous Vehicles", "Pedestrian Detection", "Neural Networks"}} and \textbf{\textit{"Safety Case"}} highlights the increasing importance of safety assurance for AI-enabled systems (e.g. ML-enabled systems). Over the past few years, especially with the advent of autonomous vehicles, there has been a growing recognition of the need to assure the safety of ML-enabled systems, and this trend is expected to grow further in the future. As ML-based systems gain momentum, assuring their safety becomes crucial. The uncertain and dynamic nature of the environments in which these systems operate introduces potentially risky unforeseen situations. Thus, it is crucial to establish mechanisms to ensure that the critical decisions made by these systems are reliable, adhere to safety standards, and can be thoroughly evaluated for potential risks. Therefore, the demand for safety assurance in ML-enabled systems is of utmost importance to build trust in these systems and mitigate the potential hazards they may face at runtime.

The map highlights the significance of keywords such as \textbf{\textit{"Regulatory Compliance" and " Safety Certification"}} in the context of safety assurance. These two concepts are closely intertwined, as certification of safety-critical systems is typically supported by using SACs to demonstrate compliance with established standards. These SACs are usually submitted to certification bodies \cite{b64}. Co-incidentally, the map also depicts keywords such as \textbf{\textit{"IEC 61508", "ARP 4754-A", "ISO 26262", "EN 5012X", "DO-178C"}} which are safety standards used in domains such as Aerospace, Automotive, and Railways. Such standards usually mandate the construction of SACs to demonstrate a system meets its requirements \cite{b64, b62}. Standard compliance and certification are critical in ensuring the safety and reliability of safety-critical systems. Failure to certify such systems may have deadly consequences. This was the case with the Titan submersible, where the company OceanGate failed to adhere to safety standards and obtain proper certification. This eventually led to the deadly implosion of that underwater vessel.

Figure \ref{fig5} also highlights the importance of approaching safety assurance through the "\textit{ \textbf{model-driven engineering}}" lens. As depicted in the figure, numerous metamodels, languages, and notations have emerged to provide formal frameworks for representing SACs. Among these, two prominent contenders are the \textbf{\textit{"GSN"}} and the \textbf{\textit{Structured Assurance Case Metamodel (SACM)}}. GSN, which has enjoyed widespread adoption for several years, has historically been the go-to choice for illustrating SACs. However, in recent times, an increasing number of researchers have transitioned to SACM, an OMG-standardized metamodel for assurance cases \cite{b78, b94}. This shift is attributed to SACM's unique qualities, as articulated in Foster et al. \cite{b62}: "\textit{SACM not only unifies and extends various predecessor notations, including GSN and CAE but also positions itself as a definitive reference model in the field of safety assurance}".

Traditional SACs are static i.e., only suitable for certifying systems before they are deployed. SACs may become obsolete during system operation \cite{b79, b80}. The presence of keywords such as \textbf{\textit{"Dynamic Safety Case"}}  and \textbf{\textit{"Change Impact Analysis"}} in Figure \ref{fig5} highlights the need for an adaptable safety assurance process allowing to monitor, assess, and adapt safety measures as a system evolves or its operational contexts change.  The need for dynamic assurance is further compounded by the \textbf{\textit{"Uncertainty"}} safety-critical systems face at runtime. That uncertainty is usually quantified by relying on assessment techniques like  \textbf{\textit{"Bayesian Networks"}}, \textbf{\textit{"Evidential Reasoning"}}, and  \textbf{\textit{"Dempster-Shafer Theory"}}.

The keywords \textbf{\textit{"Automotive", "Autonomous Vehicles"}} both indicate that safety assurance plays a crucial role in the automotive domain, particularly in the context of autonomous vehicles. Such vehicles hold the promise of revolutionizing transportation by offering increased convenience, improved efficiency, and enhanced safety. So, assuring the safety of these vehicles is of paramount importance, as they operate without direct human intervention and rely on complex algorithms and sensors to make critical decisions on the road. The need to assure such vehicles has translated in a surge of primary studies.
Their line of research  aligns with the business needs of several companies (e.g., Tesla,
Ford, Waymo, and General Motors) that are racing to become leaders in the production of autonomous driving technologies. For instance, Tesla's Robotaxi service \cite{b51} and General Motors' Cruise \cite{b52} emphasize safety in autonomous vehicle development, reflecting the importance of safety assurance in the development of their products.


Keywords such as \textbf{\textit{"Aerospace", "UAS"}} and \textbf{\textit{"UAV"}} indicate the importance of safety assurance in the field of aerospace. In research, ensuring the safety of Unmanned Aerial Vehicles (UAVs) has been a significant focus \cite{b6, b48, b49, b50}. This translated in a quite high number of studies in that domain. The growing use of UAVs in applications like surveillance, delivery, and agriculture highlights the need for robust safety assurance. That line of research aligns with the business needs of companies like Amazon, that needs to assure the safety of its devices (e.g., Prime Air delivery drones \cite{b53}).


On the map (Figure \ref{fig5}), keywords like \textbf{\textit{"Medical", "Railway"}},  \textbf{\textit{"Nuclear"}}, and \textbf{\textit{"Offshore pipeline"}} indicate that the SAC trend is also gaining momentum in health (e.g., applications in infusion pumps) , railway (e.g., autonomous trains), maritime (e.g., autonomous ships) and nuclear domains (e.g., nuclear power plants). 


\subsubsection{Temporal evolution of most influential keywords:}

\begin{figure*}[h]
  \centering
  \includegraphics[width=0.95\linewidth]{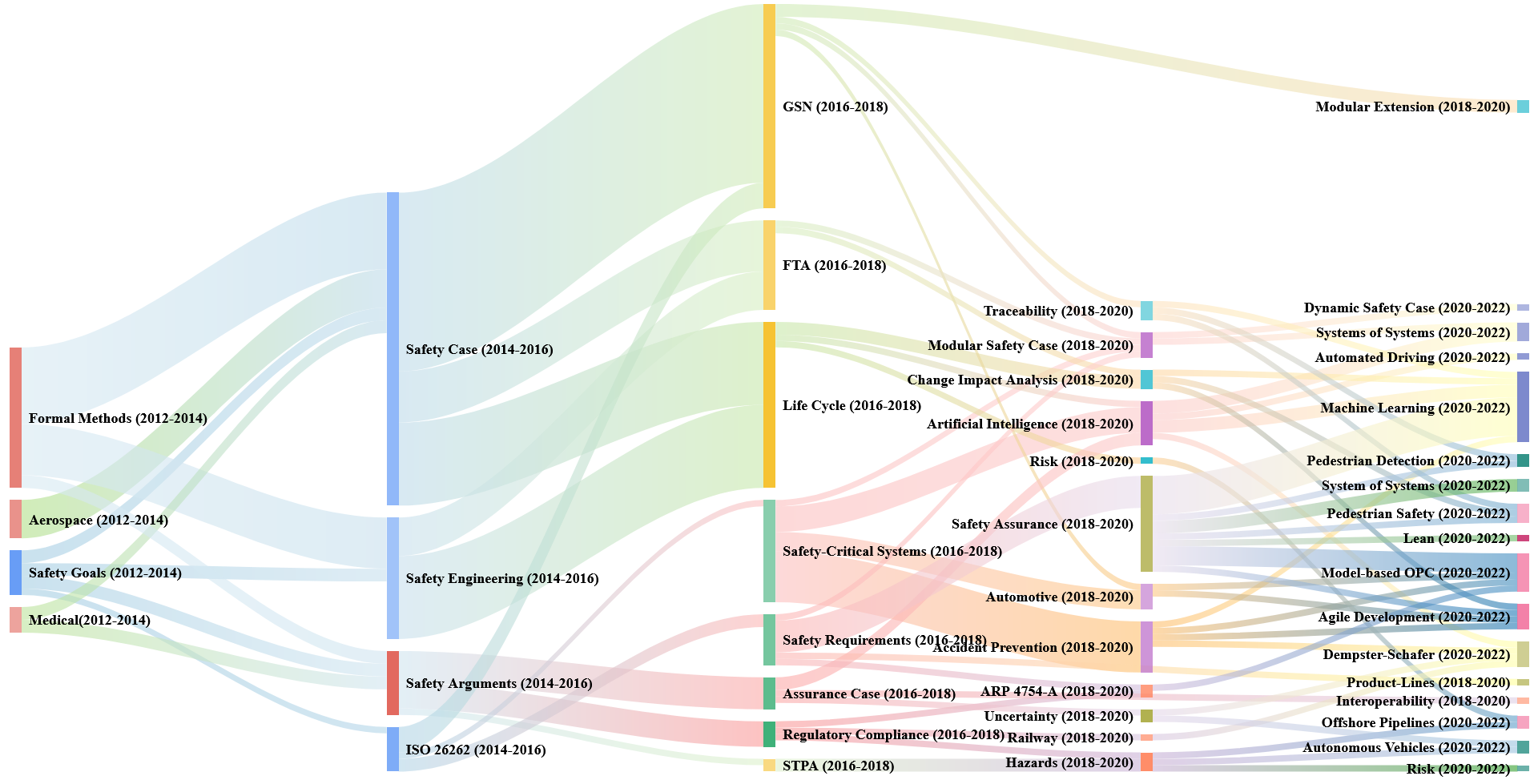}
  \caption{Temporal evolution of the most influential keywords from 2012-2022.}
  \Description{This Sankey diagram was generated with the help of Google charts \cite{b82}. It shows the evolution of the keywords depicted in \ref{fig4} from 2012 to 2022.}
  \label{fig10}
\end{figure*}

The Sankey diagram that Figure \ref{fig10} depicts, shows the temporal evolution of the most influential keywords depicted in Figure \ref{fig5}. We generated this diagram by using Google Charts \cite{b82}. Each block within the Sankey diagram represents the most frequently occurring keyword and its associated sub-period.\footnote{The connections between blocks and their respective thicknesses allow quantifying the interrelationships between different keywords across various time periods. A thicker connecting line indicates a stronger association between the two concepts \cite{b3}}. 

The Sankey diagram in Figure \ref{fig10} offers insights into the evolution of SACs. The application domains of SACs have evolved, with initial use (2010s) in aerospace and medical fields expanding to include automotive, railway, and, more recently, the oil engineering sector, with a significant focus on autonomous vehicle technologies. In the early 2010s, static SACs were developed using \textbf{\textit{formal methods (2010s)}}. At some point, the focus of these methods has shifted mainly to notations such as \textbf{\textit{GSN (2016s)}}. Over the time, other categories of SACs have emerged. First, \textbf{\textit{Modular safety cases}} emphasizing modularity. Then, in the years 2020-2022, \textbf{\textit{Dynamic safety cases}} and \textbf{\textit{Agile safety cases}} have emerged to cope with the complexity and heterogeneity of safety-critical systems (e.g., CPSs). Agile safety cases adopt \textbf{\textit{Agile development}} and \textbf{\textit{Lean}} practices.

The Sankey diagram also sheds light on the temporal evolution of safety assurance, in the context of standards and their adoption in specific domains. Around the year 2014, there was a notable surge in the attention given to \textbf{\textit{ISO 26262}}. This standard, primarily focused on functional safety in road vehicles, gained prominence as SACs found their way into the automotive industry. ISO 26262 played a pivotal role in setting safety benchmarks and guidelines for this domain. In the subsequent years, the emphasis on regulatory compliance became increasingly pronounced. This trend indicates a growing recognition of the importance of adhering to industry-specific regulations and standards to ensure safety. Compliance with these standards is crucial, particularly in safety-critical domains such as automotive and aerospace. The keyword "\textbf{\textit{automotive}}" gained momentum in the years following regulatory compliance, indicating a continued focus on safety within the automotive sector. This likely reflects the ongoing integration of safety measures, including the use of SACs, during the development and operation of vehicles. Additionally, the emergence of a new standard, \textbf{\textit{ARP 4754-A}}, specifically designed for civil aviation, underscores the need for domain-specific safety standards. This standard likely addresses the unique challenges and safety requirements within the aerospace domain, further emphasizing the importance of tailored safety measures in different industries. 

Figure \ref{fig10} also illuminates the emergence of AI and ML within the safety assurance landscape, primarily driven by the uncertainty inherent to the unpredictability of the operational contexts in which CPSs operate. Notably, all three keywords—\textbf{\textit{AI, ML, and CPSs}}—are present within the 2020-2022 timeframe, signifying not only a current upward trajectory but also a strong indication of their anticipated prominence in the future research.

\subsubsection{Characteristics of primary studies by venues}
\begin{figure}[h]
  \centering
  \includegraphics[width=\linewidth]{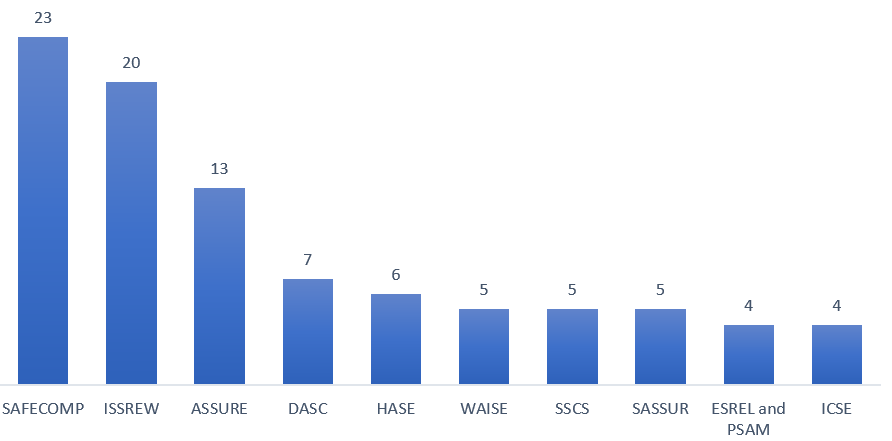}
  \caption{Bar chart showing the top 10 primary venues.}
  \Description{This bar chart, generated by Microsoft Excel, enlists the top 10 primary venues of the 224 primary studies included in our analysis.}
  \label{fig6}
\end{figure}

The bar chart in Figure \ref{fig6} depicts the top 10 venues of the primary studies included in our bibliometric analysis. The SAFECOMP Conference (International Conference on Computer Safety, Reliability, and Security) emerges as the leading venue, featuring 23 studies. The rationale is that SAFECOMP has an exclusive dedication to safety assurance. By focusing solely on that aspect, SAFECOMP provides a specialized platform for researchers, practitioners, and experts to share their insights and advancements in the field of safety assurance. Following closely is ISSREW (International Symposium on Software Reliability Engineering Workshop), with 20 studies, and ASSURE (International Workshop on Assurance Cases for Software-Intensive Systems) securing the third position with 13 publications. This trend indicates that a significant number of studies on SACs are published in conferences and workshops that focus on safety assurance. It is worth noting that workshops such as ASSURE, WAISE (International Workshop on Artificial Intelligence Safety Engineering), and SASSUR (International Workshop on Next Generation of System Assurance Approaches for Critical Systems) are conducted as integral components of the SAFECOMP conference. This observation reinforces the significance of SAFECOMP as a key conference for the dissemination of research on safety assurance. 

\subsubsection{Characteristics of primary studies by citations in Google Scholar}

Similar to Deng et al. \cite{b20}, we also conducted a bibliometric analysis to identify the most cited primary studies. Table \ref{tab:cite}\footnote{We created Table \ref{tab:cite}, based on 219 studies, as five of the studies (i.e., \cite{b22}, \cite{b23}, \cite{b24}, \cite{b25}, \cite{b26}) were not accessible on Google Scholar at the time we searched them. } presents the 15 most cited studies based on their number of citations.

\begin{table*}
  \caption{Top 15 most cited primary studies from Google Scholar as of September 23, 2023.}
  \label{tab:cite}
  \begin{tabular}{p{1cm} p{3cm} p{11cm} p{1.5cm}}
    \toprule
    No & Author & Title & \# Citations\\
    \midrule
    1&  Denney et al. (\citeyear{b27}) & Tool support for assurance case development & 107 \\ 
    1& Birch et al. (\citeyear{b21}) & Safety cases and their role in ISO 26262 functional safety assessment & 107 \\ 
      
      3&  Denney et al. (\citeyear{b5}) & AdvoCATE: An assurance case automation toolset & 104 \\ 
      4&  Denney et al. (\citeyear{b28}) & Dynamic Safety Cases for Through-Life Safety Assurance & 98 \\ 
      5&  Denney et al. (\citeyear{b29}) & A formal basis for safety case patterns & 68 \\ 
      6&  Yamamoto et al. (\citeyear{b30}) & An evaluation of argument patterns to reduce pitfalls of applying assurance case & 61 \\ 
       7& Gallina (\citeyear{b31}) & A model-driven safety certification method for process compliance & 59 \\ 
       8& Nair et al. (\citeyear{b32}) & An evidential reasoning approach for assessing confidence in safety evidence & 55 \\ 
       9& Dardar et al. (\citeyear{b33}) & Industrial experiences of building a safety case in compliance with ISO 26262 & 51 \\ 
       10& Denney et al. (\citeyear{b34}) & A lightweight methodology for safety case assembly & 48 \\ 
       11& Denney et al. (\citeyear{b35}) & Automating the assembly of aviation safety cases & 48 \\ 
       12& Ayoub et al. (\citeyear{b36}) & A systematic approach to justifying sufficient confidence in software safety arguments & 46 \\ 
       13& Weinstock et al. (\citeyear{b37}) & Measuring assurance case confidence using Baconian probabilities & 45 \\ 
        14 &Ayoub et al. (\citeyear{b38}) & A safety case pattern for model-based development approach & 44 \\ 
        15 &Guiochet et al. (\citeyear{b39}) & A model for safety case confidence assessment & 43 \\ 
  \bottomrule
\end{tabular}

\end{table*}

Two of these studies (i.e. \cite{b27} and \cite{b21}) jointly scored the highest number of citations (i.e. 107). Denney et al. are the authors of \cite{b27}. That study introduces AdvoCATE, a tool designed to automate the creation and management of safety assurance arguments, with a specific focus on unmanned aircraft systems (UAS). 
Birch et al. are the authors of \cite{b21}. That study  explores the role of SACs in the functional assessment process defined by the ISO 26262 standard.  

Interestingly, the studies that Table \ref{tab:cite} reports mostly focus on  critical topics such as the development of SACs, their assessment, and the use of patterns to foster the reuse of SAC argument structures. They also focus on  dynamic safety assurance, the use of SACs in compliance with industrial standards (e.g., ISO 26262), and the development of tools automating the safety assurance process.


\subsection{RQ2: Diversity in the SAC Literature}

\begin{figure}[h]
  \centering
  \includegraphics[width=\linewidth]{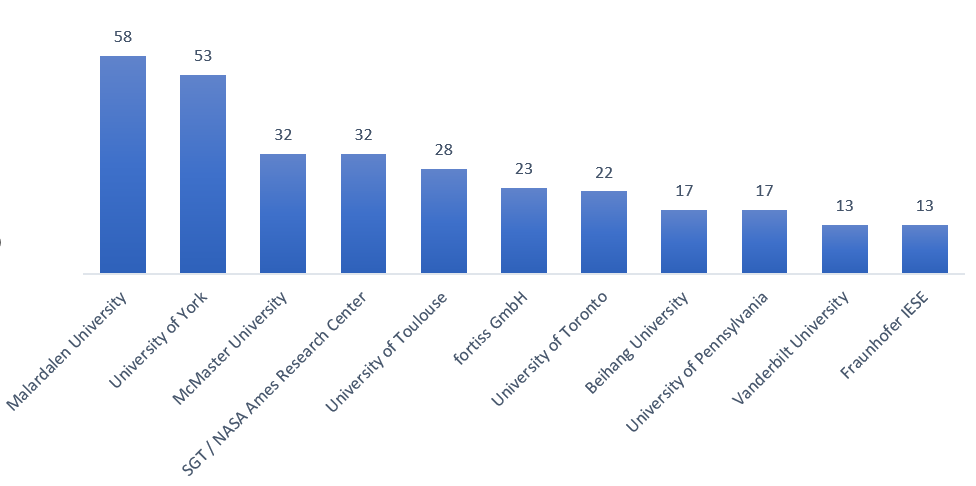}
  \caption{Bar chart showing the top affiliations of authors.}
  \Description{This bar chart, generated by Microsoft Excel, enlists the top affiliations among all the authors of the 224 primary studies.}
  \label{fig7}
\end{figure}

\subsubsection{Characteristics of primary studies by author's affiliation:}

Figure \ref{fig7} displays the top organizations and universities based on the number of authors contributing to the field of safety assurance, among the 224 primary studies included in this study. Notably, Malardalen University from Sweden stands out as the leading institution with 58 authors affiliated with that university. The University of York (United Kingdom) follows closely in second place with 53 authors, while McMaster University (Canada) and SGT/NASA Ames Research Center (USA) share the third position with 32 researchers affiliated with these organizations. The University of Toulouse arrives at the fifth position of that ranking with 28 authors.

It is interesting to analyze the findings of this chart in relation to what Figure \ref{fig8} depicts. In Figure \ref{fig8}, the USA ranked first with authors contributing 155 times, indicating a significant overall contribution from various American organizations. However, when considering the specific organizations, we observe that no single affiliation from the USA secures the top position in Figure \ref{fig7}. This is due to the heterogeneous nature of contributions from multiple organizations in that country. On the other hand, Malardalen University's first-place position in Figure \ref{fig7} can be attributed to the concentrated efforts of the university itself, as it emerges as a prominent contributor in the field of safety assurance.
Such insights shed light on the diverse and multifaceted nature of safety assurance research and the varying roles played by different organizations in advancing the field.

Interestingly, thanks to Figure \ref{fig7}, researchers can pinpoint organizations that have shown a significant commitment and expertise in safety assurance, and identify potential collaboration opportunities with experts affiliated with these institutions. 

\subsubsection{Characteristics of primary studies by author's country of affiliation:}

\begin{figure}[h]
  \centering
  \includegraphics[width=\linewidth]{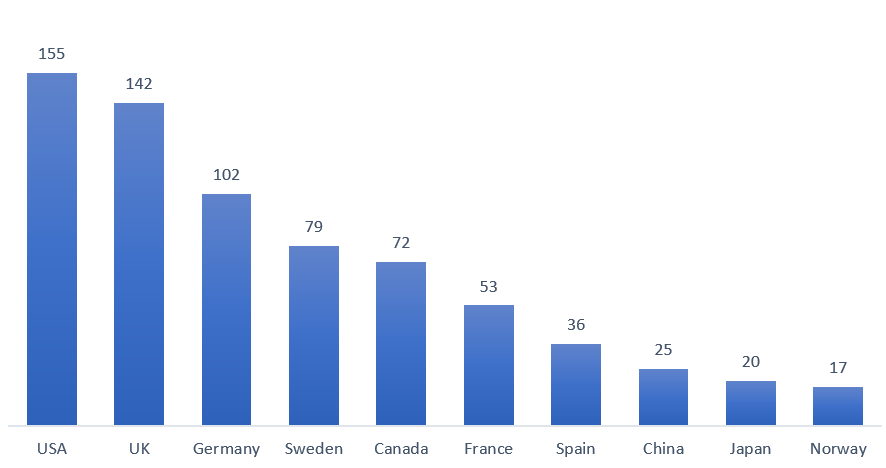}
  \caption{Bar chart showing the top countries of affiliations.}
  \Description{This bar chart, generated by Microsoft Excel, enlists the top 10 countries of affiliations among all the authors of the 224 primary studies.}
  \label{fig8}
\end{figure}

Figure \ref{fig8} illustrates the top 10 countries of author's affiliations across the 224 studies included in our analysis. The United States (USA) emerges as the top country, with 155 authors who have contributed to SAC literature during the past decade. The United Kingdom (UK) arrives at the second position, with 142 authors. Germany arrives at the third position, with 102 authors.

The prominence of the USA, UK, and Germany may be attributed to the fact that these countries have established numerous standards such as ISO 26262 \cite{b54}, DO-178C \cite{b55}, and EUROCAE \cite{b56}, which mandate the development of SACs for a given system. Consequently, authors from these countries contribute extensively to the topic due to the prevalence of safety standards and processes in their respective countries. Furthermore, the high occurrence of safety-related disasters (e.g., Columbia shuttle disaster, accidents caused by autonomous driving cars, the recent Titan submersible disaster), in some of these countries may also contribute to the increased focus on developing state-of-the-art techniques on SACs. These countries recognize the critical need for advancing safety practices and, as a result, attract significant contributions from authors in the field. With the volume of contributions depicted in Figure \ref{fig8}, the top leading nations could become the go-to nations for other nations who may be interested in formulating SAC policies.

Figure \ref{fig8} reveals notable differences in the contribution of authors from various countries to the field of SACs. For instance, countries such as Japan and Norway are shown to have relatively fewer authors who have actively contributed to this domain. Several interpretations can be made regarding this observation. For example, in Japan, safety disasters such as the Fukushima Daiichi nuclear accident may have underscored the critical need for robust national and international safety standards and guidelines. While the number of studies with respect to SACs might not be as high as its other counterparts such as Sweden, France, China, etc., the country has implemented alternative safety measures and standards to prevent and address such safety disasters in the future \cite{b90}. 

Similarly, in the case of Norway, while SACs may not be widely practiced, the country has a strong focus on safety measures. Norway has stringent regulations and safety standards to ensure the protection of its natural environment and the safety of workers in these sectors \cite{b91}. The emphasis on safety and the presence of well-established safety protocols demonstrate a different approach to safety assurance compared to safety cases. 

To complement the bar chart (Figure \ref{fig8}), we used Tableau \cite{b12} to create a world map that shows the top 15 countries with the highest number of contributions from authors in Figure \ref{fig9}. That Figure is in the Appendix.


Figures \ref{fig8} and \ref{fig9} both illustrate a pattern of moderate diversity within the landscape of SAC literature. More specifically, the majority of the literature originates from two continents, namely North America and Europe, with their respective institutions significantly represented in this domain. North America, particularly the USA and Canada, emerges as a focal point for SAC research. 
Interestingly, Figures \ref{fig8} and \ref{fig9}, also underscore China's growing influence and contribution to the field of safety assurance in Asia.


Still, some countries from Asia, Oceania, South America and Africa are not explicitly depicted in Figures \ref{fig8} and \ref{fig9}. While some of these countries play significant roles in the global economy, their absence may be attributed to several factors. First, authors with nationalities of these countries and who are actively involved in SAC research may be affiliated with institutions or organizations in foreign countries, such as the USA, UK, or Canada. Therefore, their contributions might be listed under these foreign countries. Second, safety assurance practices can vary based on regional priorities and industries. Addressing this moderate diversity in SACs literature requires fostering global collaboration, and encouraging research initiatives in under-represented regions.

\subsection{RQ3: Future research directions}
Our bibliometric analysis findings show several research directions still need to be explored to yield efficient safety assurance solutions:

\subsubsection{Develop advanced solutions to foster regulatory safety compliance}

Figure \ref{fig5} highlights that the majority of primary studies in our bibliometric analysis mainly focus on safety assurance standards in the Automotive (e.g., ISO 26262), Aerospace (e.g., ARP 4754-A and DO-178C) and Railway (e.g., EN 501X) domains. However, a notable gap exists in specific standards that explicitly mandate safety assurance for modern AI-enabled CPSs. This gap is concerning given the increasing prevalence of AI in safety-critical domains. Collaboration between AI companies and governments, as Sam Altman suggests \cite{b69}, could lead to the creation of a regulatory agency to oversee AI system development and safety testing. Implementing AI regulations poses challenges such as balancing national and global standards and addressing tech giants' concerns about staying competitive. Standards like UL 4600 \cite{b92} for autonomous vehicles and safety cases are essential for ensuring compliance and enhancing accountability in AI-enabled systems.

\subsubsection{Integrating Safety Assurance into Agile/DevOps: Aligning Safety Case Development with Agile Methodologies}
Figure \ref{fig5} shows that there has been relatively limited research dedicated to agile safety cases. The rationale for this observation could be attributed to the fact that agile safety cases are considered an emerging field within the broader domain of safety assurance. The concept of agile safety cases is relatively new and may not have gained as much traction as more traditional approaches in safety case development. However, in recent years, as shown in Figure \ref{fig10}, there has been a noticeable increase in the adoption of agile development methods for safety-critical software. This shift has been driven by the desire to streamline the software development process, reduce costs, and improve overall software quality.
Embracing an agile approach allows safety cases to be developed incrementally as information becomes available during the project, leading to greater safety awareness and understanding throughout the development life cycle \cite{b71, b72, b73}.

\subsubsection{Dynamic Assurance Approaches for Cyber-Physical Systems (CPSs): From Development to Life Cycle Management}
Emerging technologies, such as CPSs (e.g., ML-enabled autonomous driving systems, and unmanned aerial systems), call for a paradigm shift in safety assurance. More specifically, our analysis from Figure \ref{fig5} allows us to conclude that traditional approaches to safety certification and assurance are not sufficient to address the dynamic nature of these systems. There is therefore a pressing need for a new class of safety certification techniques that continuously assess and evolve safety reasoning, aligning with the system's life cycle. This transformative approach aims to provide through-life safety assurance, encompassing not only the initial development and deployment stages but also run-time monitoring based on real-world operational data \cite{b28}. To ensure safety and reliability in modern complex systems, dynamic safety assurance, as a continuous and adaptable process, becomes essential. This area holds significant promise for future research and development in safety assurance

Neto et al. 
\cite{b43} further echo the need to support the dynamic safety assurance of intelligent safety-critical systems.


\section{Threats to Validity} \label{sec6}

Based on the framework that Wohlin et al. \cite{b66} and Zhou et al. \cite{b67} proposed, we have identified the following threats to validity.

\subsection{Internal validity}

Several bibliometric analyses select studies based on the first three phases of our selection strategy (see Section \ref{sec39}). This usually yields several hundred/thousands of studies and makes the selection process less time-consuming. This may also yield a relatively high amount of noise in the so-selected studies. We, therefore, built on the features of common secondary studies guidelines (e.g., PRISMA 2020 and SEGRESS) to propose a more rigorous selection strategy that consists of five phases. This had the advantage of significantly reducing the noise in the selected studies by allowing the selection of studies that are more relevant to the research topic. However, this also significantly reduced the number of primary studies included in our bibliometric analysis and therefore reduced the scope of our bibliometric analysis. Nonetheless, our bibliometric analysis analyses several hundred of studies and can therefore serve as a reference to position future work on the safety assurance topic.

\subsection{Conclusion validity}
The field of safety assurance is inherently interdisciplinary, demanding expertise in many domains including those we explored in our analysis. A lack of such interdisciplinary expertise could result in an incomplete analysis that may not provide enough insights into the surveyed topic. To mitigate that issue, our research team is constituted with members with strong expertise in computer science (with a focus on software engineering and machine learning), along with extensive experience in automotive and aerospace. 

\section{Conclusion} \label{sec7}

Our bibliometric analysis harnesses common bibliometric tools, to quantitatively analyze primary studies drawn from well-known databases and focusing on safety assurance cases (SACs). This allows examining the trends in publication years, and the top venues where the primary studies were published. This also allows exploring the most influential keywords in the field of SACs, the most cited primary studies, and the diversity of affiliations of authors. This allows identifying future directions in the SAC research.

As future work, we aim at building upon the foundations laid by this study by shifting our focus towards a qualitative synthesis of the data extracted from primary studies. This will allow getting additional insights on the topic under investigation.

\bibliographystyle{ACM-Reference-Format}
\bibliography{sigconf-ref}

\appendix

\section{Appendix}

\begin{figure}[h]
  \centering
  \includegraphics[width=\linewidth]{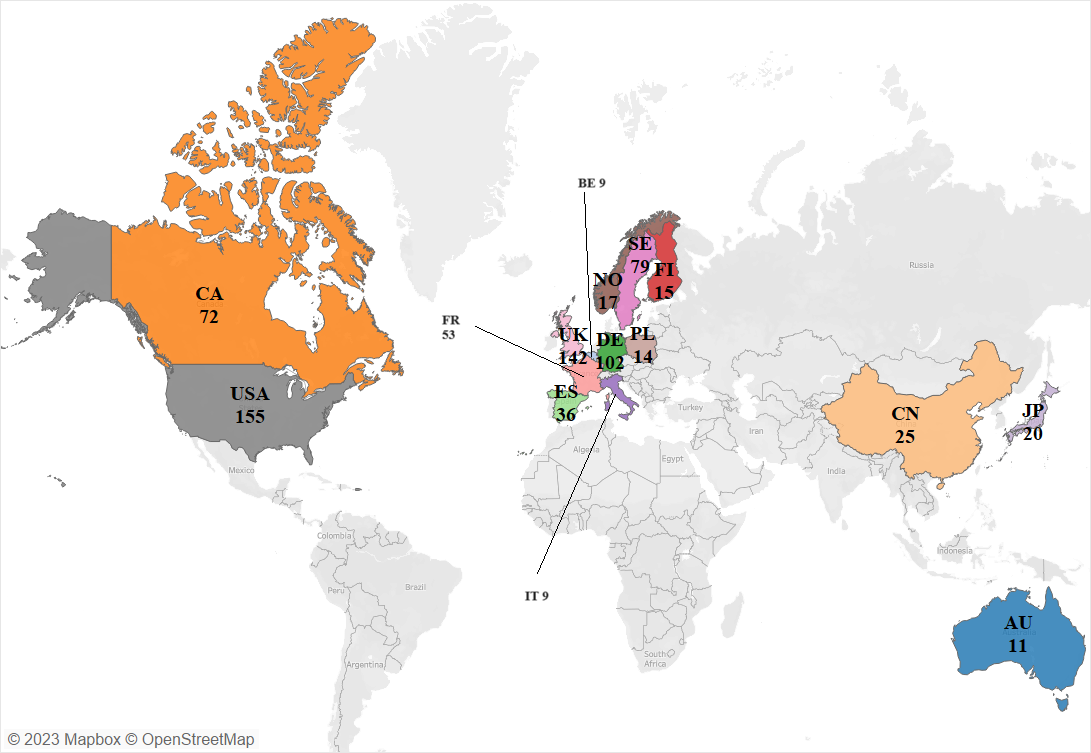}
  \caption{World Map showing the top 15 countries.}
  \Description{This world map, generated with the help of Tableau showcases the top 15 countries with most number of authors. The countries here refer to the country of the author's affiliation.}
  \label{fig9}
\end{figure}

The world map in Figure \ref{fig9} shows  top 15 countries of affiliation\footnote{In that Figure, USA = United States of America; UK = United Kingdom; DE = Germany; SE= Sweden; CA = Canada; FR= France; ES= Spain; CN= China; JP = Japan; NO = Norway; FI = Finland; PL= Poland; AU= Australia; IT = Italy; BE= Belgium}.

\end{document}